# Observing bulk diamond spin coherence in high-purity nanodiamonds

Helena S. Knowles[1†*], Dhiren M. Kara[1†] and Mete Atatüre[1*]

[1]*Cavendish Laboratory, University of Cambridge, JJ Thomson Avenue, Cambridge CB3 0HE, UK*

[†]*These authors contributed equally to this work.*

[*]*Corresponding authors: hsk35@cam.ac.uk and ma424@cam.ac.uk*

**Nitrogen-vacancy centres (NVs) in diamond are attractive for research straddling quantum information science[1-8] and nanoscale magnetometry[9-13] and thermometry[14,15]. While ultrapure bulk diamond NVs sustain the longest spin coherence times among optically accessible spins[16-18], nanodiamond NVs display persistently poor spin coherence[17,19,20]. Here we introduce high-purity nanodiamonds accommodating record-long NV coherence times, >60 $\mu$s, observed via universal dynamical decoupling[21]. We show that the main contribution to decoherence comes from nearby nitrogen impurities rather than surface states. We protect the NV spin free precession, essential to DC magnetometry, by driving solely these impurities into the motional narrowing regime. This extends the NV free induction decay time from 440 ns, longer than that in type Ib bulk diamond[22], to 1.27 $\mu$s, which is comparable to that in type IIa (impurity-free) diamond[23]. These properties allow the simultaneous exploitation of both high sensitivity and nanometre resolution in diamond-based emergent quantum technologies[24].**

NVs are stable, fluorescent point defects in diamond consisting of a substitutional nitrogen (N) atom and a neighbouring vacancy site (Fig. 1a). They have a spin triplet ground state sensitive to magnetic and electric fields. The optically detected magnetic resonance (ODMR) technique[25] demonstrates their optical and microwave (MW) addressability, where NV fluorescence is reduced when MW field is applied on resonance with a spin transition. A typical ODMR measurement is shown in Fig. 1a, where the two allowed spin transitions ($m_s$ =

0 → $m_s$ = -1 and $m_s$ = 0 → $m_s$ = +1) are split by 270 MHz under a static magnetic field, $B_{ext}$, of strength 4.8 mT.

NV spin coherence in bulk diamond is affected by its interaction with nearby electronic and nuclear spins, referred to as the spin reservoir. Typically, this reservoir comprises electronic spins of substitutional N atoms and nuclear spins of the $^{13}$C isotope (1.1 % abundant) in the lattice. In type Ib (N < 500 ppm) and isotopically engineered type IIa (N < 0.1ppm and $^{13}$C < 0.3 %) bulk diamond NV coherence times of 100 $\mu s^{26}$ and 3 ms[18] have been reached, respectively, by employing dynamical decoupling schemes to suppress the reservoir effects. In contrast, nanodiamond NVs display disappointingly short coherence times in the range of 1.4 to 10.9 $\mu s^{17,19,20}$, limiting their use in quantum technologies. Although the exact cause for the reduction of NV spin coherence is unclear, it has been attributed tentatively to the insuppressibly fast dynamics of the nanodiamond surface spins and the typically high N concentration inside the nanodiamonds. While suppressing the former decoherence mechanism through complete surface-state passivation remains a challenge[24], the latter can in principle be avoided by using cleaner diamond as the raw material. However, the use of high-purity type IIa diamond can be ruled out due to the scarcity of NVs in this category. Further, milling NV-implanted bulk diamond films would also provide only a modest yield of NV-containing nanodiamonds.

In this work we introduce nanodiamonds milled from synthetic high-pressure high-temperature (HPHT) bulk diamond with a nominal impurity concentration (<50 ppm) that is intermediate to type Ib and type IIa categories to approach an optimal condition for both nanodiamond purity and NV yield without requiring irradiation or advanced nanofabrication processes. Figure 1b is a high resolution atomic force micrograph of the nanodiamonds deposited on quartz and the histogram in Fig. 1c reveals that their diameters are predominantly within 10-35 nm (23 ± 7 nm). On average a few per cent of these nanodiamonds contain NVs. To investigate their coherence properties we perform spin-echo (SE) measurements on several single-NV-containing nanodiamonds using a home-built confocal microscope setup with a MW transmission wire (see Methods and Supplementary Information). The experimental sequence shown in the inset of Fig. 1d consists of optical initialization of the NV spin into the $m_s$=0 state ($|0\rangle$), application of MW pulses in the SE sequence, and then readout of the |0> population, $P_{|0>}$, as a function of the total duration, $\tau$. As $\tau$ increases, spin decoherence leads to a drop in $P_{|0>}$. Figure 1d presents SE measurements for eleven NVs. The solid and dashed curves are fits of the theoretically expected $\exp[-(\tau/T_{SE})^\alpha]$ form, where $T_{SE}$ is the characteristic spin coherence time. In the fast decoherence limit, where the spin reservoir evolves on timescales shorter than the inter-pulse separation, the exponent $\alpha$ is 1[27]. All previous reports on nanodiamond samples are in this limit showing clear exponential decay[17,19,20]. In contrast, the histogram in Fig. 1d shows that $\alpha$, extracted from the fits to data in Fig. 1c, is between 2 and 2.7. This is clear evidence of a quasi-static reservoir, where the theoretically predicted range for $\alpha$ is between 2 and 3[28].

Universal dynamical decoupling can be used to prolong the coherence time of a spin only if it is embedded in a quasi-static reservoir[22]. We apply this technique to four NVs, identified in Fig. 1d with solid symbols: NV1 (blue, $T_{SE}$ = 5.8 µs), NV2 (brown, $T_{SE}$ = 5.1 µs), NV3 (purple, $T_{SE}$ = 4.8 µs) and NV4 (green, $T_{SE}$ = 5.8 µs), replacing the single refocussing π-pulse in the SE sequence with a train of $n_\pi$ π-pulses, as illustrated at the top of Fig. 2. Figure 2a displays a set of measurements on NV1 for $n_\pi$ ranging from 1 (solid red circles) to 71 (solid blue circles). The curves are fit to $\exp[-(\tau/T_{DD})^2]$, where $T_{DD}$ is the extended spin coherence time. For $n_\pi$ = 71 we reach $T_{DD}$ of 67 µs exceeding not only the highest reported nanodiamond coherence, but also that in type Ib bulk diamond (<200 ppm N concentration) for the same $n_\pi$[22]. We note that this is achieved using a pulse sequence that does not correct for pulse errors, unlike the CPMG sequence[29].

The theoretical enhancement factor, $T_{DD}/T_{SE}$, in a truly quasi-static reservoir scales as $[n_\pi^{-2/3} + (T_{SE}/T_1)]^{-1}$ for a finite NV spin lifetime, $T_1$ [22,27]. The dashed blue curve in Fig. 2b is $T_{DD}/T_{SE}$ for infinite $T_1$ and the dashed red curve is for $T_1 = 30 T_{SE}$. The experimental $T_{DD}/T_{SE}$ scaling for NV1-NV3 lie within the shaded region, consistent with their independently measured $T_1$'s (typically between 150 $\mu$s and 1.3 ms) demonstrating the universality of this scaling (examples of shorter $T_1$ are provided in the Supplementary Information). The inset of Fig. 2b shows that the experimental $T_{DD}$ values for NV1 follow the theoretical prediction (solid blue curve) based on the measured values of $T_1 = 1.25$ ms and $T_{SE} = 5.8$ $\mu$s. Moreover, the lack of any observable deviation in enhancement indicates that $T_{DD}$ of 67 $\mu$s is not bounded and can be extended further with increasing $n_\pi$. As a counterexample, the modest enhancement of $T_{DD}/T_{SE}$ for NV4 (green solid triangles) deviates from the expected scaling with its $T_1 = 459$ $\mu$s and $T_{SE} = 5.8$ $\mu$s – possibly due to close proximity of an N-N impurity pair. Nevertheless, NV4 coherence reaches $T_{DD} = 20.8$ $\mu$s for $n_\pi = 21$, still longer than any previous report. Overall, the SE and DD data suggest in unison that the dominant reservoir in our nanodiamonds is the quasi-static paramagnetic impurities within the crystal, *e.g.* N atoms, rather than the rapidly evolving surface states.

We now operate the NV as a magnetometer to probe the properties of this spin reservoir: Monitoring the NV SE coherence at $\tau = 2$ $\mu$s whilst scanning the frequency of a radiowave (RF) pulse applied simultaneously with the MW $\pi$-pulse (see top right of Fig. 3) maps the spin reservoir resonances. The electron spin of N couples anisotropically to its nuclear spin due to the Jahn-Teller distortion resulting in four quasi-static (on ms timescale) groups, $N_K = N_a$, $N_b$, $N_c$ and $N_d$, determined by the diamond crystal axes illustrated in Fig. 3a (see Supplementary Information). Previous measurements have shown that strong $B_{ext}$ parallel to the NV axis renders the transitions of $N_K$ groups and any purely electronic paramagnetic centres spectrally indistinguishable[22]. We work in a weak $B_{ext}$ of 5.4 mT (comparable to the average hyperfine coupling strength) aligned 10° from the NV symmetry axis to resolve $N_K$ groups and other species (*e.g.* surface dangling bonds). The blue points in Fig. 3b display the reservoir spectrum for NV2 measured with a weak 400-ns RF pulse. A theoretical model based on the $N_K$-Hamiltonian and the RF-induced manipulation of the reservoir (see Methods and Supplementary Information 4) predicts the spectrum for each $N_K$ group, as shown in Fig. 3c. The solid red curve in Fig. 3b is the direct sum of these four contributions and matches the complex measured spectrum with remarkable consistency.

The linewidth of these resonances, $\Gamma$, is determined by the N concentration, $C_N$, in the weak excitation limit via $\Gamma = (0.00205\ C_N^2 + 0.239)^{\frac{1}{2}}$, where $C_N$ is in ppm and $\Gamma$ is in MHz[30]. We select the isolated 147.8-MHz transition (red arrow in Fig. 3a) originating from the $N_a$ group. A high resolution measurement and a Lorentzian fit (red curve) are shown in Fig. 3d. The extracted $\Gamma$ of 1.7 ± 0.2 MHz corresponds to $C_N < 36 ± 5$ ppm – consistent with that of the synthetic diamond (<50 ppm) from which the nanodiamonds were milled. This is an order of magnitude lower than previously studied nanodiamonds (200 ppm[19] and 500-1000 ppm[20]). Finally, at this magnetic field, purely electronic spin impurities would have resonance at 152 MHz in Fig. 3d (vertical dashed black line). The absence of signal at this frequency rules out contributions to NV decoherence from such impurities.

Dynamical decoupling is an invaluable tool for measuring fast AC fields with a sensitivity level given by $\delta B_{ac} \sim \pi h\ [2\ g\ \mu_B\ C\ (T_{DD})^{\frac{1}{2}}]^{-1}$, where $h$ is the Planck constant, g is the spin g-factor and $C$ is a constant describing the measurement efficiency[9]. For typical experimental conditions ($C \sim 0.05$) our nanodiamonds offer at least $\delta B_{ac}$ = 140 nT/Hz$^{\frac{1}{2}}$. However, many applications depend on detecting DC or slowly evolving fields, such as variations in magnetic landscape within living cells and Ramsey interferometry becomes instrumental instead. Here, a spin phase acquired by the NV during a free precession time, $\tau$, is monitored to detect magnetic fields. Without the benefit of decoupling the NV from the reservoir via refocussing π-pulses on the NV spin, an alternative approach, realized in pure bulk diamond only[22], is to drive the reservoir electromagnetically, so that its dynamics occurs at a rate much faster than its coupling to the NV. This motional averaging regime is reached by satisfying the condition, $\Omega_N / \gamma \gg 1$, where $\Omega_N$ and $\gamma$ denote the driven precession rate of the reservoir spins and the NV-reservoir dipolar coupling strength, respectively. Figure 4a displays reservoir spin Rabi oscillations in our nanodiamonds obtained by recording the NV SE signal as a function of RF-pulse duration, $T_{RF}$, when driving the $N_a$ group at 147.8 MHz. The period of the oscillation reveals $\Omega_N$ = 11.5 MHz. To obtain a value for $\gamma$ we use the spin echo double resonance (SEDOR) protocol, where a standard SE sequence is synchronized with a RF π-pulse applied to the $N_a$ group (top right of Fig. 4). Figure 4b demonstrates that when this particular dephasing channel is turned on, the SEDOR signal decays faster than the SE signal. Fitting the SEDOR measurement with $\exp[-(\gamma\tau)^2/2]$ yields a $\gamma$ of 0.66 ± 0.03 MHz. Thus, $\Omega_N / \gamma$ = 17.4 ($\gg$ 1) confirms that our nanodiamonds can accommodate the motionally averaged reservoir regime.

The orange data in Fig 4c show spin coherence of NV2 as a function of free evolution time, *i.e.* Ramsey interference, and the inset illustrates the sequence. The beating arises from the hyperfine level splitting in NV2 (2.2 MHz) and oscillations from an intentional detuning (16 MHz) of the MW from the spin transition. A free evolution spin coherence time, $T_2^*$, of 0.44 ± 0.01 μs is extracted. This value is longer than $T_2^*$ for NVs in type Ib bulk diamond (<200 ppm N concentration) – as expected from the higher purity material we use. The green data in Fig. 4c show Ramsey Double Resonance (RADOR) – Ramsey interferometry with five additional RF frequencies driving every $N_K$ group to satisfy ($\Omega_N / \gamma$) > 10 (see Supplementary S4.4). In this motionally averaged reservoir $T_2^*$ gains an almost 3-fold prolongation reaching 1.27 ± 0.04 μs. This timescale indicates that we are able to recover 75% of the $T_2^*$ in impurity-free type IIa bulk diamond[23] (<0.1 ppm N concentration) where the reservoir comprises only $^{13}C$ nuclear spins.

Our results confirm indisputably that obtaining nanodiamonds from synthetic diamond containing N impurity concentration that is intermediate to type Ib and type IIa categories results in spin coherence well beyond what has been attainable to-date, while maintaining a feasible yield of NVs. Under typical operational conditions our nanodiamonds offer a sensitivity of 0.6 $\mu$T/Hz$^{1/2}$ for DC (and 140 nT/Hz$^{1/2}$ for AC) magnetometry. As a nanoscale thermometer our nanodiamonds could sense 3.5 mK temperature change within 1 minute. In parallel, by employing electron-nuclear double resonance protocols in the motionally-averaged reservoir regime for N spins, utilizing nearby $^{13}$C nuclei as quantum registers[2] may be possible in nanodiamonds. Finally, this impurity concentration corresponds to less than 40 N spins in a 20-nm diameter nanodiamond (< 5 for 10 nm) and should allow the detection of correlated dynamics such as spin squeezing[31] within this truly mesoscopic spin reservoir.

## Methods

**Setup:** All experiments were carried out on non-detonation nanodiamonds, milled from synthetic HPHT bulk diamond containing less than 50 ppm N impurities (Nabond) and then drop-cast onto a quartz plate. To optically address and detect individual NVs a scanning confocal microscope setup with an air-based 0.9 NA objective (Nikon 100x) was used. Non-resonant laser light (532 nm, Laser Quantum Ventus VIS 532) was used to excite the NVs and the resulting fluorescence was detected on an avalanche photodiode (PerkinElmer). An amplified MW signal (Rohde&Schwarz SMF100A and 16 W amplifier, Mini-Circuits) was coupled through a 20-$\mu$m diameter copper wire attached to the sample surface. A PulseBlaster ESR-PRO card and switches from Mini-Circuits allowed the synchronisation of the pulse sequences to nanosecond precision. Laser control was achieved by AOM switching in a single pass setup (see Supplementary Information for more details).

**Spin reservoir model**: Each N group is described by the following Hamiltonian:

$$\hat{H} = a_{\parallel}\hat{S}_z\hat{I}_z + a_{\perp}(\hat{S}_x\hat{I}_x + \hat{S}_y\hat{I}_y) - q\hat{I}_z^2 \qquad (1)$$

where $a_{\parallel}$ = 114.2 MHz and $a_{\perp}$ = 81.8 MHz are the parallel and perpendicular hyperfine constants with respect to the Jahn-Teller axis defined as the *z*-direction, *q* = 3.97 MHz is the quadrupole interaction, and the electron and nuclear spin is *S*=1/2 and *I*=1 respectively.

**RADOR measurement**: We simulate the RADOR signal of Fig. 4 using a function of the form:

$$S_{\text{RADOR}} = e^{-(t/T_2^*)^2} \sum_{i=1}^{3} R_i \cos[\omega_i t + \varphi] . \qquad (2)$$

The values $\omega_1$, $\omega_2$ and $\omega_3$ correspond to the detuning of the MW driving frequency with each of the hyperfine levels of the NV giving rise to the observed beating signal and $R_i$ is the amplitude of each frequency component.

**Acknowledgements**

We gratefully acknowledge financial support by the University of Cambridge, the European Research Council (FP7/2007-2013)/ERC Grant agreement no. 209636, and the FP7 Marie Curie Initial Training Network S³NANO. We also thank T. Muller, S. Topliss and Y. Alaverdyan for technical assistance and F. Jelezko, R. Hanson and C. Degen for fruitful discussions.

# Figure Captions

**Figure 1 – Nanodiamond size and NV coherence statistics. a,** An illustration of a NV defect in a diamond lattice and ODMR in 4.8-mT magnetic field. **b,** Atomic force micrograph of nanodiamonds deposited on quartz. **c,** Histogram of nanodiamond diameters yielding a mean value of 23 nm. **d,** Population of the $m_s$ = 0 state after the application of a spin echo sequence, illustrated in the left inset. Data is shown for several single NVs where NV1, NV2, NV3 and NV4 are plotted in solid blue circles, purple squares, brown diamonds and green triangles, respectively. Solid and dashed lines are fits of the form $\exp[-(\tau/T_{SE})^\alpha]$, where $T_{SE}$ and $\alpha$ are free parameters. Right inset, histogram of $\alpha$ values reveal a mean of 2.3 indicating a predominantly slow, quasi-static reservoir for all NVs investigated.

**Figure 2: Universal dynamical decoupling in nanodiamonds. a,** Coherence of NV1 as pulse sequence (shown in illustration above a) is extended from $n_\pi$ = 1 (solid red circles) through $n_\pi$ = 3, 9, 21, 51 to 71 (solid blue circles) with their corresponding fits to $\exp[-(\tau/T_{DD})^2]$ leading to $T_{DD}$ reaching 67 μs. **b,** The measured $T_{DD}$ vs. $n_\pi$ for NV1 (inset) where the coherence time enhancement follows the power law $T \propto n_\pi^{2/3}$. The enhancement factor, $T_{DD}/T_{SE}$, for four NVs presented in Fig. 1d demonstrates this universal scaling. NVs 1, 2 and 3 lie within the shaded blue region identified by assuming infinite $T_1$ (upper bound) and $T_1 = 30\, T_{SE}$ (lower bound). The enhancement factor for NV4 (solid green triangles) is an example of deviation from the theoretically expected behaviour.

**Figure 3: NV Magnetometry on the spin reservoir. a,** Illustration of the NV in the diamond lattice surrounded by N atoms with four distinct orientations due to JT distortion: $N_a$ parallel to the NV axis and $N_b$, $N_c$ and $N_d$ along the other three crystal axes. **b,** Spectrum of the reservoir spins obtained by monitoring NV SE signal at $\tau=2$ μs and scanning the frequency of a RF pulse (scheme illustrated above panel b). The spectrum can be reproduced by a model of N impurity electrons anisotropically coupled to their nuclear spins. **c,** Theoretical spectrum predicted for each group for $B_{ext}$ = 4.8 mT and 10° angle with respect to the NV orientation. The sum of all four spectra is plotted as the solid red curve in panel b. The slight discrepancy in the peak amplitudes arise from the variations in RF field strength during the measurements and from the lack of knowledge of the exact sites of the N impurities (see Supplementary information). **d,** Low power spectrum of the transition marked with the red arrow in panel b. From the FWHM of 1.7 MHz we extract a nitrogen impurity concentration of < 36 ppm. The dashed line marks the resonance frequency expected for purely electronic paramagnetic centres.

**Figure 4: Coherent control and motional averaging of reservoir spins. a,** Rabi oscillations of the $N_a$ group. The transition at 148 MHz is coherently driven in the middle of a spin echo sequence, as illustrated above the panel. **b,** SEDOR on the same group shown in blue solid circles compared to the SE decay on NV without reservoir control (solid orange circles). The solid curves are fits to the corresponding data. **c,** Free evolution of the NV spin in a Ramsey interferometry measurement (upper plot) and with a sum of five CW RF signals applied during $\tau$, driving the N spins into the motional averaging limit (lower plot). Solid lines are fits that include a Gaussian coherence decay and the detuning of the driving field from the three hyperfine levels of the NV spin interacting with its host N. The coherence times of 0.44 μs (upper plot) and 1.27 μs (lower plot) are extracted from the envelope decay.

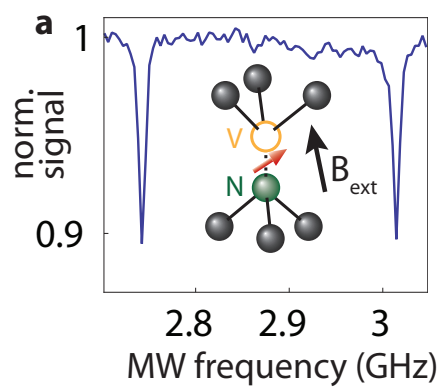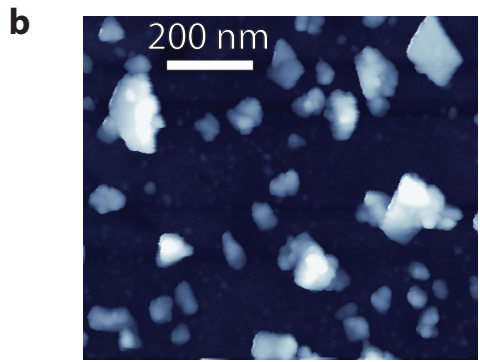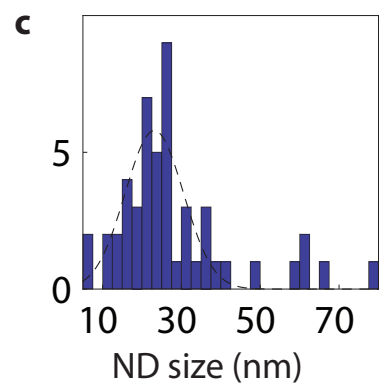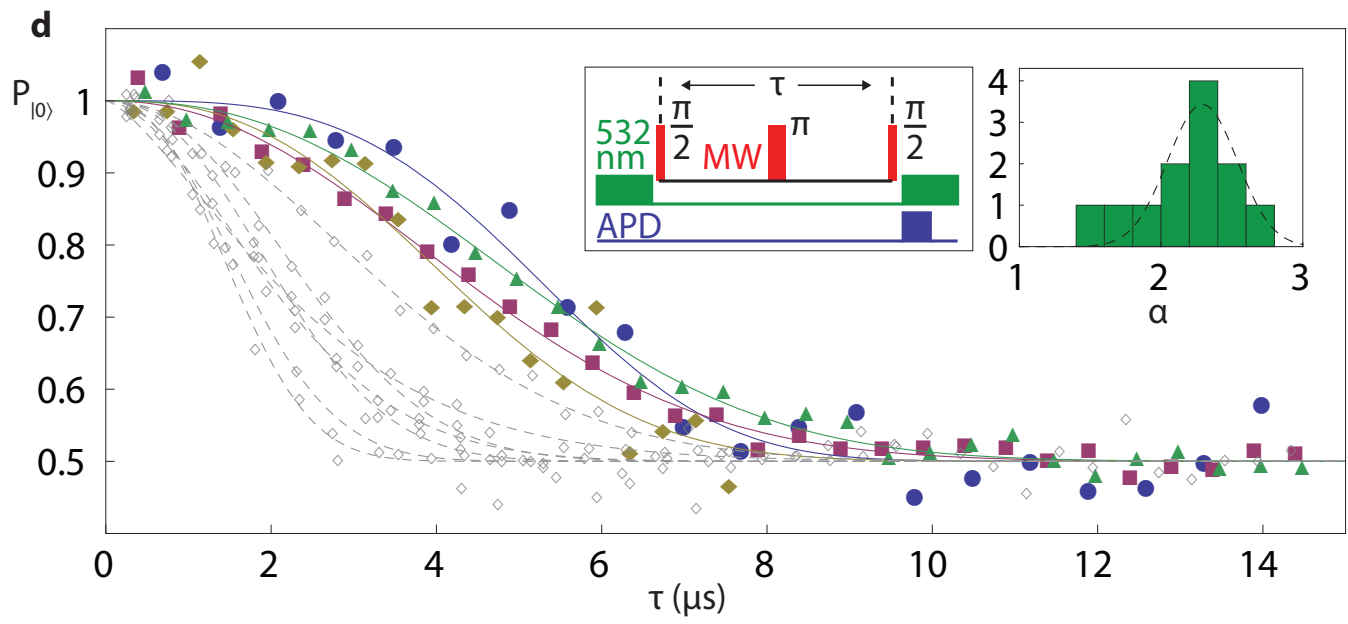

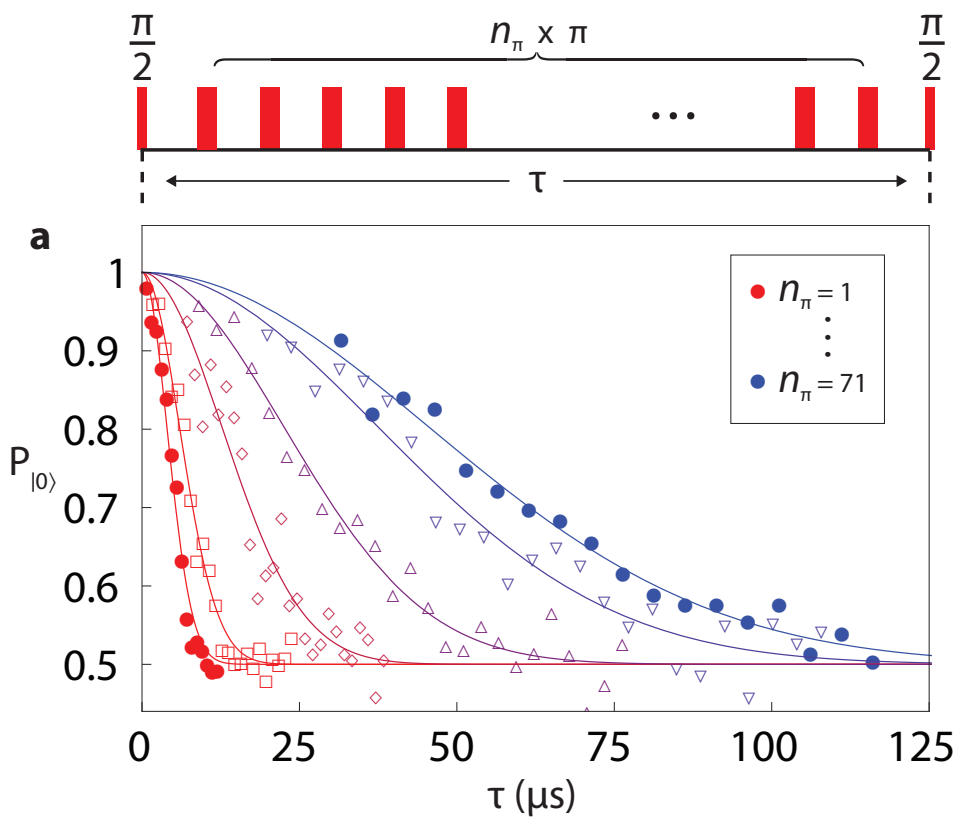

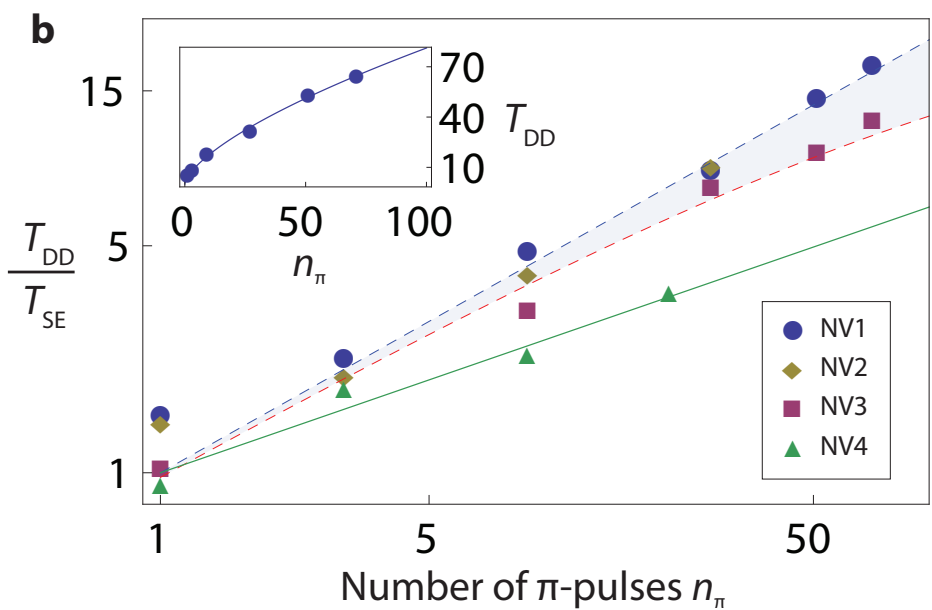

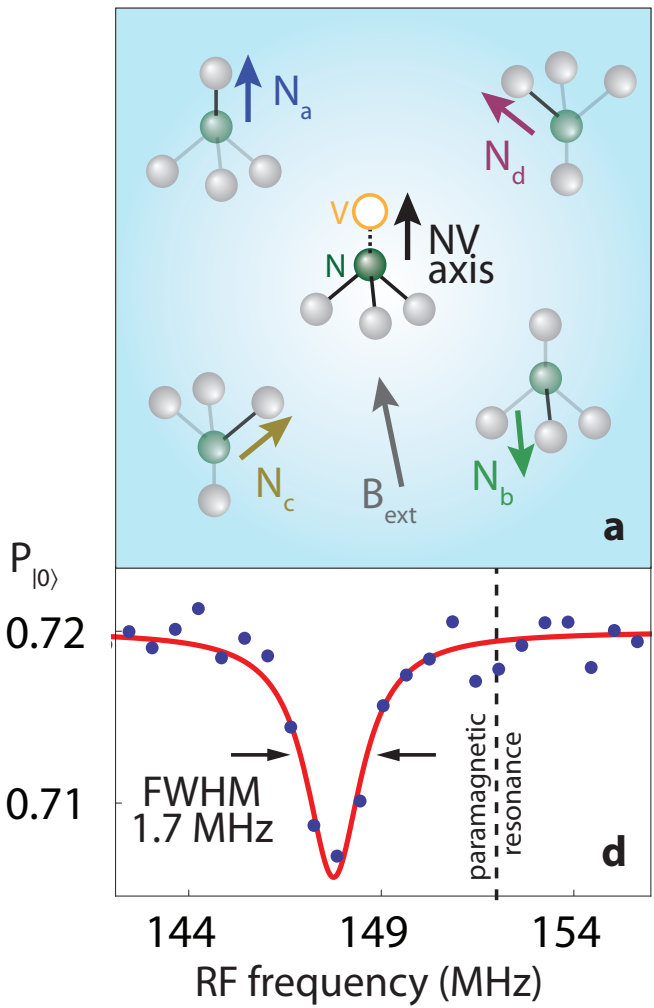
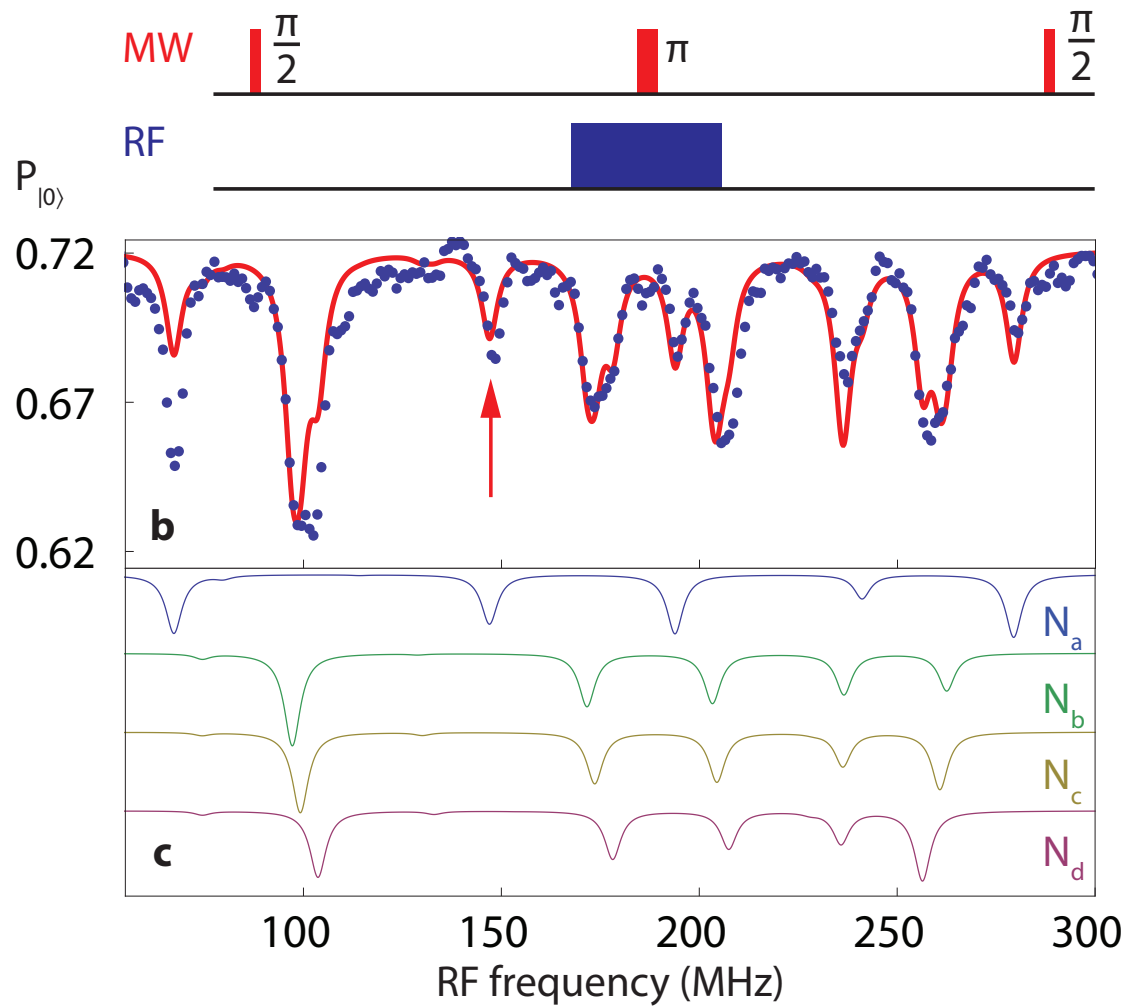

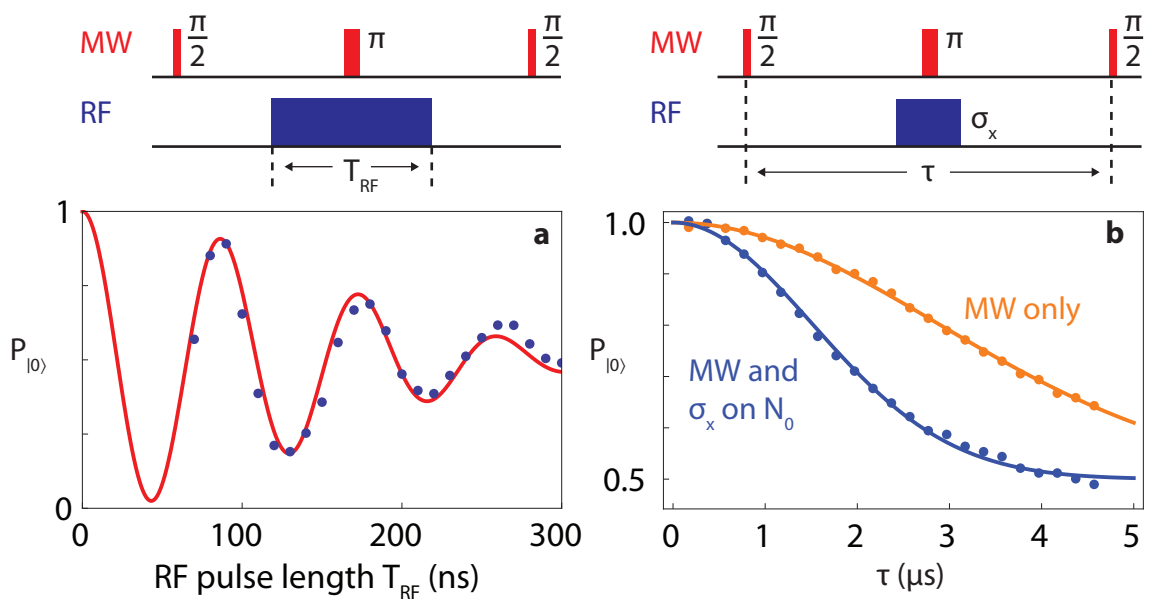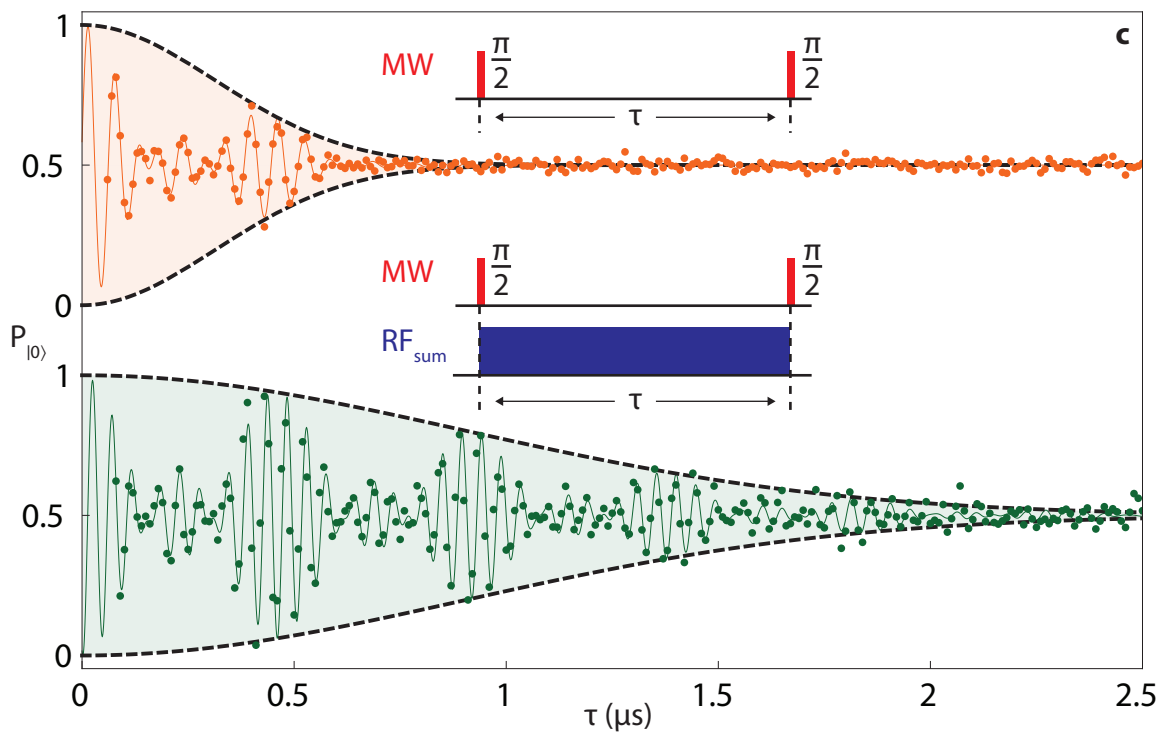